\documentclass[%
superscriptaddress,
preprint,
amsmath,amssymb,
prb
]{revtex4-1}

\usepackage{graphicx}
\usepackage{dcolumn}
\usepackage{bm}

\usepackage{color} 

\begin{document}

\title{Edge states of Floquet-Dirac semimetal in a laser-driven semiconductor quantum-well}

\author{Boyuan Zhang}
\affiliation{Doctoral Program in Materials Science, Graduate School of Pure and Applied Sciences, University of Tsukuba, Tsukuba, Ibaraki 305-8573, Japan}
\author{Nobuya Maeshima}
\affiliation{Center for Computational Sciences, University of Tsukuba, Tsukuba 305-8577, Japan}
\author{Ken-ichi Hino}
\affiliation{Division of Materials Science, Faculty of Pure and Applied Sciences, University of Tsukuba, Tsukuba 305-8573, Japan\\\vspace{0.5cm}
{\rm Correspondence and requests for
materials should be addressed to K.H. (email: hino@ims.tsukuba.ac.jp)\\\vspace{0.8cm}
}
}
\affiliation{Center for Computational Sciences, University of Tsukuba, Tsukuba 305-8577, Japan}


\begin{abstract}
Band crossings observed in a wide range of condensed matter systems are recognized as a key to understand low-energy fermionic excitations that behave as massless Dirac particles.
Despite rapid progress in this field, the exploration of non-equilibrium topological states remains scarce and it has potential ability of providing a new platform to create unexpected massless Dirac states.
Here we show that in a semiconductor quantum-well  driven by a cw-laser with  linear polarization, the optical Stark effect conducts bulk-band crossing, and the resulting Floquet-Dirac semimetallic phase supports an unconventional edge state in the projected one-dimensional Brillouin zone
under a boundary condition that an electron is confined in the direction 
perpendicular to that of the laser polarization.
Further, we reveal that this edge state mediates a transition between topological and non-topological edge states that is caused by tuning the laser intensity.
We also show that the properties of the edge states are strikingly changed under a different boundary condition.
It is found  that such difference originates from that nearly fourfold-degenerate points
 exist in a certain intermediate region of the bulk Brillouin zone between high-symmetry points.

\end{abstract}

\maketitle


The theoretical prediction and the subsequent discovery of topological insulators\cite{Kane2005, Bernevig2006}
have led to explosive expansion of the studies of topological perspectives of condensed matter\cite{Hasan2010, Qi2011} and photonic crystals,\cite{Ozawa2019} where a sharp distinction between topologically trivial and non-trivial phases with energy gaps is made by the presence of a gapless Dirac dispersion.
The viewpoint of the gapless state has been developed to connect to the studies of topological semimetals akin to graphene,\cite{Yao2009, Akhmerov2008,Castro2009} termed Dirac, Weyl, and line-node semimetals.\cite{Wehling2014, Armitage2018}
Emergence of topological gapless phases is derived from symmetries inherent in the physical system of concern, namely, the time-reversal (T-)symmetry, the spatial-inversion (I-)symmetry, small groups supported by space groups, and so on.\cite{Wehling2014,Armitage2018,Murakami2007a,Murakami2007b,Young2012,Young2015,Wang2012,Wang2013a,Yang2014,Park2017}
As regards a Dirac semimetal (DSM), this is realized by an accidental 
band crossing due to fine-tuning of material parameters,\cite{Murakami2007a,Murakami2007b}
the symmetry-enforced mechanism,\cite{Young2012,Young2015} and
the band inversion mechanism.\cite{Wang2012,Wang2013a,Yang2014}
Further, there exist edge modes known as double Fermi arcs at the surface of the DSM 
formed by the band inversion mechanism.
\cite{Yang2014,Yi2014,Xu2015,Kargarian2016}
Recently, a growing attention has been paid to two-dimensional (2D) DSMs from the perspective of in-depth theories and applications to novel nano scale devices.
\cite{Young2015,Park2017,Doh2017, Ramankutty2018,Luo2020}

While these intriguing topological semimetals are fabricated in equilibrium, 
there is still concealed attainability of creating and manipulating gapless Dirac dispersions in Floquet topological systems with spatiotemporal periodicity.
Owing to this property, the existence of quasienergy bands are ensured by the Floquet theorem.\cite{Shirley1965, Kitagawa2010}
These systems are driven into non-equilibrium states by a temporally periodic external-field that has many degrees of freedom of controlling these states in terms of built-in parameters.\cite{Oka2009, Zhenghao2011, Linder2011, Wang2013b, Rechtsman2013, Wang2014, Taguchi2016, Hubener2017, Claassen2017, 
Nakagawa2020}
It is reported that a three-dimensional (3D) DSM, ${\rm Na_3Bi}$, is changed to a Floquet-Weyl semimetal by irradiation of femtosecond laser pulses with a circularly polarized light,\cite{Hubener2017}
and that band crossings at Dirac points are realized by forming a photonic Floquet topological insulator mimicking a graphene-like honeycomb lattice driven by a circularly polarized light.\cite{Rechtsman2013}
It is remarked that the T-symmetry is broken/protected in a system under the application of a circularly/linearly polarized light-field.

In this study, first, we show that a gapless Dirac state emerges in a 2D-bulk band of a semiconductor quantum well driven by a cw-laser with a linear polarization, where the T-symmetry is protected, however, the I-symmetry is broken.
Here, the optical Stark effect (OSE) accompanying quasienergy band splitting \cite{Autler1955, Knight1980} is introduced to cause
an accidental band crossing at high-symmetry points in the 2D Brillouin zone (BZ).
This effect is enhanced by a nearly resonant optical excitation from a valence ($p$-orbital) band to a conduction ($s$-orbital).
Such an optically nonlinear excitation leads to strong hybridization between the different parity states with $s$- and $p$-orbitals over a wide range of the BZ due to the broken I-symmetry.
Second, we show that such photoinduced hybridization brings the resulting DSM state to coincide with an unconventional edge state with a linear and nodeful dispersion in a projected one-dimensional (1D) BZ under a boundary condition that an electron is confined 
in the direction perpendicular to that of the applied electric field of laser.
Further, when the laser intensity changes to make a gap open, this edge state is transformed smoothly into another edge state within this gap; which is
either topologically trivial or non-trivial.
It is also shown that the manifestation of these edge states is drastically changed under 
another boundary condition that an electron is confined in the direction parallel to the 
applied electric field.
To deepen the understanding of the properties and boundary-condition dependence 
of the edge states, 
we introduce an interband polarization function that reflects the degree of parity hybridization in the bulk BZ. 
Finally, we point out that local anticrossings with quite small energy separation exist
in a certain intermediate region of the 2D BZ between high-symmetry points, and 
show that these anticrossings leading to nearly fourfold degeneracy are
crucial to understand the different properties of edge states,
depending on the boundary conditions.

These edge states concerned here share features with other studies.
As regards the OSE, a valley-selective OSE is demonstrated in monolayer transition
metal dichalcogenides with application of a circularly polarized electric field.\cite{Sie2015}
As regards edge states of the Floquet DSM states,
Tamm states \cite{Tamm1932, Shockley1939, Ohno1990} appearing in the surfaces of several Dirac materials are theoretically examined. \cite{Volkov2016}
Recently, growing interest has been captured in the interrelation of Tamm states with topological edge states in optical waveguide arrays,\cite{Longhi2013, Wang2018, Chen2019}
1D photonic crystals,\cite{Tsurimaki2018, Lu2019, Henriques2020} a graphene ring with the Aharonov-Bohm effect,\cite{Latyshev2014} a honeycomb magnon insulator,\cite{Pantaleon2019} and a gold surface.\cite{Yan2015}

\bigskip
\noindent
{\bf Results}

\noindent
{\bf Modified Bernevig-Hughes-Zhang model with a laser-electron interaction.}
We begin by constructing the Hamiltonian of the present system of a semiconductor quantum well with a linearly polarized light field based on the paradigmatic Bernevig-Hughes-Zhang (BHZ) model\cite{Bernevig2006} composed of two bands with $s$- and $p$-orbitals in view of a spin degree of freedom. 
Hereafter, the band with $s(p)$-orbital is termed as $s(p)$-band just for the sake of simplicity, and the atomic units (a.u.) are used throughout unless otherwise stated.
The BHZ Hamiltonian concerned here is read as the $4 \times 4$-matrix:
\begin{equation}
\mathcal{H}_{\rm{BHZ}}(\boldsymbol{k})
=\epsilon(\boldsymbol{k})I+\sum_{i=3}^5 d_i(\boldsymbol{k})\Gamma_i
\label{BHZ}
\end{equation}
with $\boldsymbol{k}=(k_x, k_y)$ as a 2D Bloch momentum defined in 
the $xy$-plane normal to the direction of crystal growth of quantum well, namely, the $z$-axis. Here $I$ represents the $4\times 4$ unit matrix, and $\Gamma_j$'s represent the
four-dimensional Dirac matrices for the Clifford algebra, defined by
\(
\Gamma_1=\tau_x\otimes\sigma_x,\:
\Gamma_2=\tau_x\otimes\sigma_y,\:
\Gamma_3=\tau_x\otimes\sigma_z,\:
\Gamma_4=\tau_z\otimes I_2
\), and 
\(
\Gamma_5=\tau_y\otimes I_2
\), where
$I_2$ represents the $2\times 2$ unit matrix, 
$\tau_s$ and $\sigma_s$ with $s=x,\:y,\:z$ represent
the Pauli matrices for orbital and spin degrees of freedom, respectively, and 
the anti-commutation relation,
\(
\{\Gamma_i, \Gamma_j \}=2\delta_{ij}
\), 
is ensured.
Further, 
\(
\epsilon(\boldsymbol{k})={1\over 2}(\epsilon_s+\epsilon_p)-(t_{ss}-t_{pp})(\cos{k_x a}+\cos{k_y a})
\), 
and 
\begin{eqnarray}
\left\{
\begin{array}{l}
d_3(\boldsymbol{k})=2t_{sp}\sin{k_ya}\\
d_4(\boldsymbol{k})={1\over 2}(\epsilon_s-\epsilon_p)-(t_{ss}+t_{pp})(\cos{k_x a}+\cos{k_y a})\\
d_5(\boldsymbol{k})=2t_{sp}\sin{k_xa}
\end{array}
\right.,
\label{dk}
\end{eqnarray}
where $\epsilon_b$ and $8t_{bb}$ represent the center and width of band $b$, respectively, and $t_{bb^\prime}$ represents a hopping matrix between $b$ and $b^\prime(\not = b)$ orbitals with lattice constant $a$; after this, $a$ is set equal to unity unless otherwise stated.
Hereafter, a semiconductor quantum well of HgTe/CdTe is accounted as the object of material. It is understood that $t_{ss}=t_{pp}$ and $\epsilon_s=-\epsilon_p$.
Thus, a Fermi energy is given by $E_F=(\epsilon_s+\epsilon_p)/2=0$, and the energy gap $E_g$ at the $\Gamma$-point of the quantum well equals $2(\epsilon_s-4t_{ss})$.

An interaction of electron with a laser field is introduced into $\mathcal{H}_{\rm BHZ}(\boldsymbol{k})$
by replacing $\boldsymbol{k}$ by $\boldsymbol{K}(t)=\boldsymbol{k}+\boldsymbol{A}(t)$, followed by adding to $\mathcal{H}_{\rm BHZ}(\boldsymbol{K}(t))$ an interband dipole interaction given by
\(
v(t)=\Omega(t)\Gamma_6
\), 
where $\Gamma_6=\tau_x\otimes I_2$, and $\Omega(t)$ is a real function of time $t$, 
provided as $\Omega(t)=\boldsymbol{F}(t)\cdot\boldsymbol{X}_{sp}$.
Here an electric field of the cw-laser with a linear polarization in the $x$-direction is given by
$\boldsymbol{F}(t)=(F_x\cos{\omega t},0,0)$ with a constant amplitude $F_x$ and a frequency $\omega$, where this is related with a vector potential $\boldsymbol{A}(t)$ as 
$\boldsymbol{F}(t)=-\dot{\boldsymbol{A}}(t)$, and
$\boldsymbol{X}_{sp}=(X_{sp},0,0)$ represents a matrix element of electric dipole transition between $s$- and $p$-orbitals, independent of $\boldsymbol{k}$: $\boldsymbol{X}_{sp}=\boldsymbol{X}_{ps}^*$.
Thus, in place of $\mathcal{H}_{\rm BHZ}(\boldsymbol{k})$, the resulting expression ends with up
\begin{equation}
H(\boldsymbol{k},t)=\mathcal{H}_{\rm BHZ}(\boldsymbol{K}(t))+v(t)
\equiv \sum_{i=3}^6D_i(\boldsymbol{k},t)\Gamma_i,
\label{H}
\end{equation}
where
$D_i(\boldsymbol{k},t)=d_i(\boldsymbol{K}(t))$ for $i\not=6$, and $D_6(\boldsymbol{k},t)=\Omega(t)$ (for more detail of derivation of it, consult Supplementary Note 1).
Obviously, this ensures the temporal periodicity, $H(\boldsymbol{k},t+T)=H(\boldsymbol{k},t)$,
with $T=2\pi/\omega$, and the system of concern follows the Floquet theorem.

\bigskip
\noindent
{\bf T- and pseudo-I-symmetries.}
It is evident that 
the T- and I-symmetries are conserved in $\mathcal{H}_{\rm BHZ}(\boldsymbol{k})$,
that is,
\(
\Theta^{-1}\: \mathcal{H}_{\rm BHZ}(-\boldsymbol{k}) \Theta=\mathcal{H}_{\rm BHZ}(\boldsymbol{k})
\), and
\(
\Pi^{-1}\: \mathcal{H}_{\rm BHZ}(-\boldsymbol{k}) \Pi=\mathcal{H}_{\rm BHZ}(\boldsymbol{k})
\), 
where $\Theta$ and $\Pi$ represent the T- and I-operators, defined by
$\Theta=-iI_2\otimes \sigma_y K$ and $\Pi=\tau_z\otimes I_2$, respectively, where 
$K$ means an operation of taking complex conjugate.
Accordingly, by fine-tuning $E_g$, it is likely that an accidental band crossing occurs at a high-symmetry point 
with four-fold degeneracy. \cite{Murakami2007a}

On the other hand, as regards $H(\boldsymbol{k},t)$, while the T-symmetry is still respected, 
the I-symmetry is broken because 
$D_i(-\boldsymbol{k},t) \not=-D_i(\boldsymbol{k},t)$ for $i=5,6$, and $D_4(-\boldsymbol{k},t) \not=D_4(\boldsymbol{k},t)$.
That is, 
\(
\Theta^{-1} H(-\boldsymbol{k},-t) \Theta=H(\boldsymbol{k}, t)
\), whereas
\(
\Pi^{-1} H(-\boldsymbol{k},t) \Pi\not =H(\boldsymbol{k},t)
\). 
In fact, it is shown that in terms of an operator defined as 
$\tilde{\Pi}=\Pi \mathcal{T}_{1/2}$, the symmetry
\(
\tilde{\Pi}^{-1} H(-\boldsymbol{k},t+T/2) \tilde{\Pi}=H(\boldsymbol{k},t)
\)
is retrieved,
where $\mathcal{T}_{1/2}$ represents the operation of putting $t$ ahead by a half period $T/2$,
namely, the replacement of $t \rightarrow t+T/2$.
Hereafter $\tilde{\Pi}$ is termed as 
the pseudo-I operator reminiscent of the time-glide symmetry.
\cite{Morimoto2017}

\bigskip
\noindent
{\bf Floquet quasienergy bands.}
Owing to the Floquet theorem, a wavefunction of the time-dependent Schr$\ddot{\rm o}$dinger equation for $H(\boldsymbol{k},t)$ is expressed as
$\Psi_{\boldsymbol{k}\alpha}(t)e^{-iE_\alpha(\boldsymbol{k})t}$ for Floquet state $\alpha$, and thus $\Psi_{\boldsymbol{k}\alpha}(t)$ is ensured by the quasi-stationary equation 
\begin{equation}
L(\boldsymbol{k},t)\Psi_{\boldsymbol{k}\alpha}(t)=E_\alpha(\boldsymbol{k})\Psi_{\boldsymbol{k}\alpha}(t)
\label{Psi}
\end{equation}
under a temporally periodic condition
$\Psi_{\boldsymbol{k}\alpha}(t+T)=\Psi_{\boldsymbol{k}\alpha}(t)$,
where
$L(\boldsymbol{k},t)=H(\boldsymbol{k},t)-iI\partial/\partial t$ and $E_\alpha(\boldsymbol{k})$ is an eigenvalue termed as quasienergy of the 2D bulk band.
It is noted that $\Theta^{-1} L(-\boldsymbol{k},-t)\Theta=L(\boldsymbol{k},t)$, and
$\tilde{\Pi}^{-1} L(-\boldsymbol{k},t+T/2)\tilde{\Pi}=L(\boldsymbol{k},t)$.
The state $\alpha$ is denoted as a combination of $\beta(n)$, where $\beta$ is assigned to either $s$- or $p$-band that dominates over this hybridized state, and $n$ represents an additional quantum number due to the temporal periodicity that
means the number of dressing photons.
Owing to the pseudo-I-symmetry,
$E_{\alpha}(\boldsymbol{k})$ equals $E_{\alpha}(-\boldsymbol{k})$, 
where the associated eigenstate of the former is $\Psi_{\boldsymbol{k}\alpha}(t)$, while that of the latter is $\tilde{\Pi}\Psi_{\boldsymbol{k}\alpha}(t)=\Psi_{-\boldsymbol{k}\alpha}(t+T/2)$.
It is remarked that a parity is still a good quantum number
at a high-symmetry point
$\boldsymbol{k}=\boldsymbol{k}^j\:\:(j=\Gamma,X_1,X_2,M)$, that is, 
$\Pi^{-1} L(\boldsymbol{k}^j,t)\Pi=L(\boldsymbol{k}^j,t)$, where
four $X$-points in the 2D-BZ are not equivalent
because of the application of the laser field in the $x$-direction,
and these are distinguished by representing as $X_1$ and $X_2$.

$E_{\alpha}(\boldsymbol{k})$'s are obtained by numerically solving Eq.~(\ref{Psi}) in the frequency $(\omega)$ domain, where the Floquet matrix is recast into
\(
\tilde{L}_{nn^\prime}(\boldsymbol{k},\omega)=( n|L(\boldsymbol{k},t)|n^\prime)
\) 
with respect to $n$ and $n^\prime$ photon states; it is understood that
\(
(n|\cdots|n^\prime)={1\over T}\int_{0}^T \:dt e^{-i(n-n^\prime)\omega t}
\cdots
\).
The matrix element of it is read as 
\begin{equation}
\tilde{L}_{nn^\prime}(\boldsymbol{k},\omega)=n\omega\delta_{nn^\prime}I
+\sum_{i=3}^6
\tilde{D}_{i,nn^\prime}(\boldsymbol{k},\omega)\Gamma_i,
\label{tildeL}
\end{equation}
where
\(
\tilde{D}_{i,nn^\prime}(\boldsymbol{k},\omega)=( n|D_i(\boldsymbol{k},t)|n^\prime)
\), 
and an explicit expression of it is given in Supplementary Note 2.
A quasienergy band,
$\mathcal{E}_{\alpha}(k_x)/\mathcal{E}_{\alpha}(k_y)$, which is the projection of $E_{\alpha}(\boldsymbol{k})$ onto the $k_x/k_y$-direction, 
is obtained by solving the equation provided by representing Eq.~(\ref{Psi}) in the lattice representation just in the $y/x$-direction where the motion of electron is confined.
Thus, there are two types of vanishing boundary conditions
that the electron is confined in the direction either perpendicular or parallel to
the direction of $\boldsymbol{F}(t)$.
Hereafter, the former type is termed the boundary condition A, the latter is the boundary condition B; the allocation of both types is schematically shown in Supplementary Figure 1.

\begin{figure}[h]
\begin{center}
\includegraphics[width=12.0cm,clip]{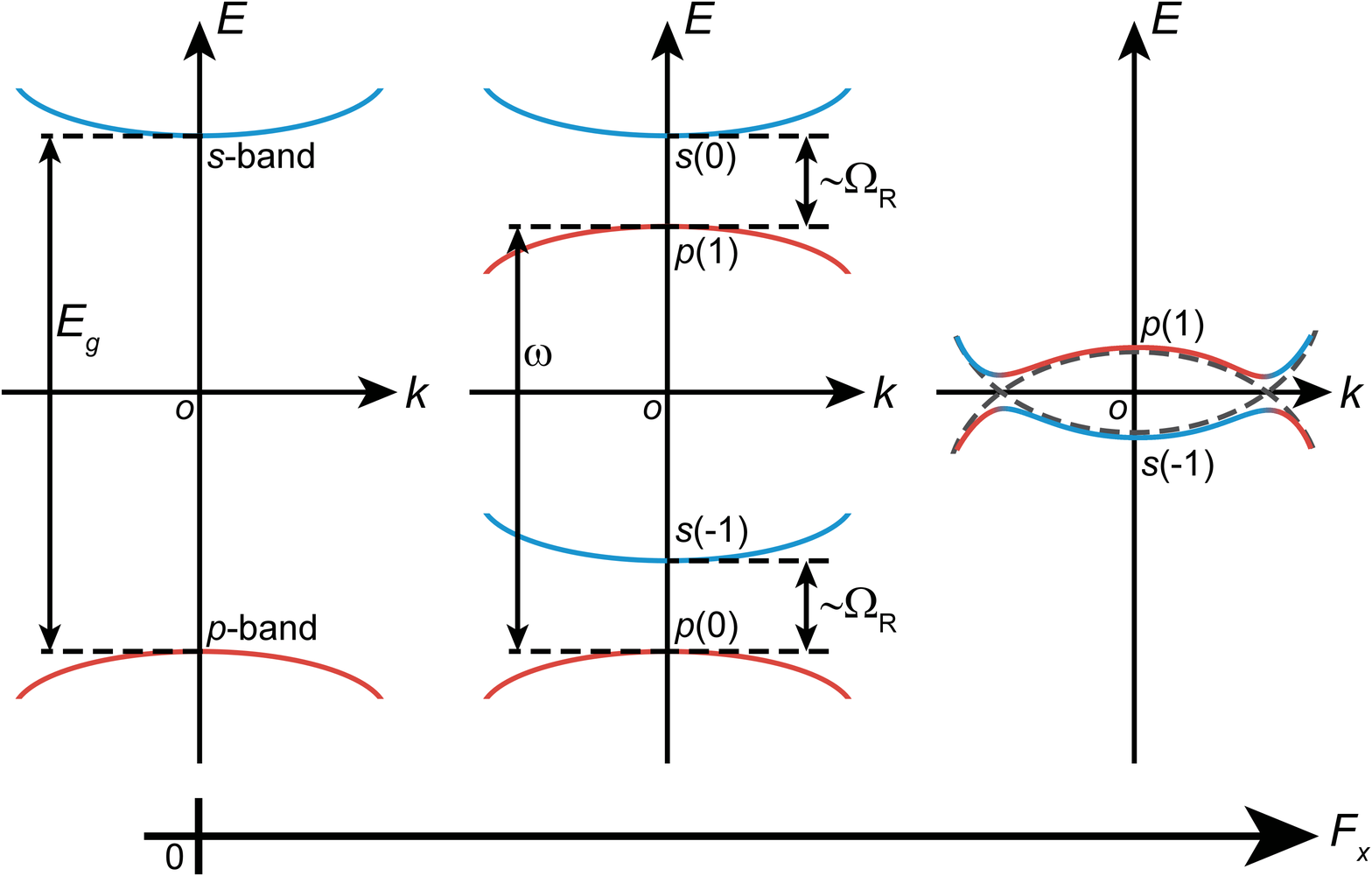}
\caption{{\bf Scheme of the nearly resonant optical-excitation followed by the OSE.}
(Left) The original energy allocation of the $p$-band (red solid line) and the $s$-band (blue solid line) with energy gap $E_g$.
(Center) With the application of cw-laser with frequency $\omega$ and constant electric 
field $F_x$, the OSE causes quasienergy-splitting of the order of the Rabi frequency $\Omega_R$ between a pair of photodressed bands, $s(n-1)$ and $p(n)$, with $n=0,1$.
(Right) With the further increase in $F_x$, a pair of bands of $p(1)$ and $s(-1)$ undergoes inversion with anticrossing. Band crossing takes place at a certain $F_x$, as shown 
by a dashed line.
}
\label{fig1}
\end{center}
\end{figure}

\bigskip\noindent
{\bf Quasienergy-band inversion and crossing due to OSE.}
Here we show an overall change of quasienergy spectra with respect to $F_x$ due to the OSE, eventually leading to a band inversion.
Figure~\ref{fig1} shows the scheme of the nearly resonant optical-excitation from the $p$-band to $s$-band with $\omega \lessapprox E_g $.
Such a scheme of excitation almost maximizes the degree of the $sp$ hybridization to
induce sharp quasienergy-splitting of the order of $\Omega_R$ between two quasienergy bands of $s(n-1)$ and $p(n)$ for $n=0,1$, where $\Omega_R$ represents the Rabi frequency given by $F_xX_{sp}$.\cite{Autler1955}
As $F_x$ increases, a pair of photodressed bands of $p(1)$ and $s(-1)$ undergoes inversion to swerve with anticrossing.

\begin{figure}[h]
\begin{center}
\includegraphics[width=15.0cm,clip]{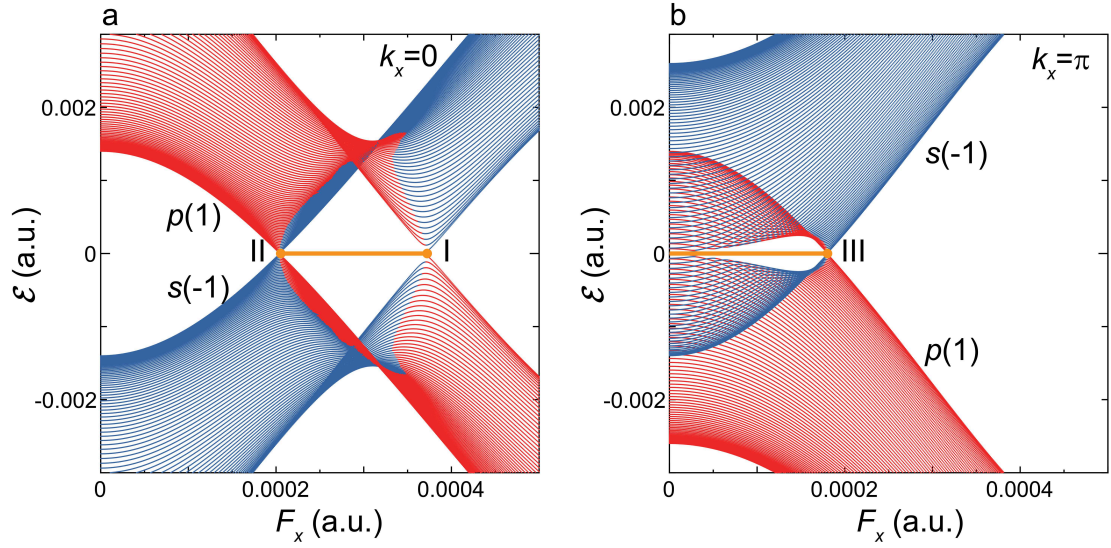}
\caption{{\bf Band inversion and band crossing.}
{\bf (a)} Shown are $\mathcal{E}_{p(1)}(k_x)$ and $\mathcal{E}_{s(-1)}(k_x)$ for $k_x=0$ as a function of $F_x$.
The two quasienergy bands $p(1)$ and $s(-1)$ (shown by red and blue lines, respectively) cross when $F_x$ is fine-tuned at the positions of I and II.
Shown are the zero modes (Dirac nodes) by a yellow solid line.
{\bf (b)} The same as the panel (a) but for $k_x=\pi$. The two quasienergy bands cross when $F_x$ is fine-tuned at the position of III.
}
\label{fig2}
\end{center}
\end{figure}

Figures~\ref{fig2}a and \ref{fig2}b show the calculated results of $\mathcal{E}_{p(1)}(k_x)$ and $\mathcal{E}_{s(-1)}(k_x)$ under the boundary condition A as a function of $F_x$ for $k_x=0$ and $\pi$, respectively.
It is noted that these bands cross at the abscissa $(\mathcal{E}(k_x)=E_F=0)$ without anticrossings at $F_x$'s indicated by I, II, and III; these positions are mentioned as $F_x^{\rm I}$, $F_x^{\rm II}$, and $F_x^{\rm III}$, respectively.
The band inversions of $p(1)$ and $s(-1)$ discerned in Figs.~\ref{fig2}a and \ref{fig2}b 
accompany the emergence of zero-energy modes indicative of topological phase transitions, where the zero-energy modes are designated by the horizontal lines along the abscissa in $F_x^{\rm II} < F_x < F_x^{\rm I}$ and $F_x < F_x^{\rm III}$, respectively.

\begin{figure}[h]
\begin{center}
\includegraphics[width=12.0cm,clip]{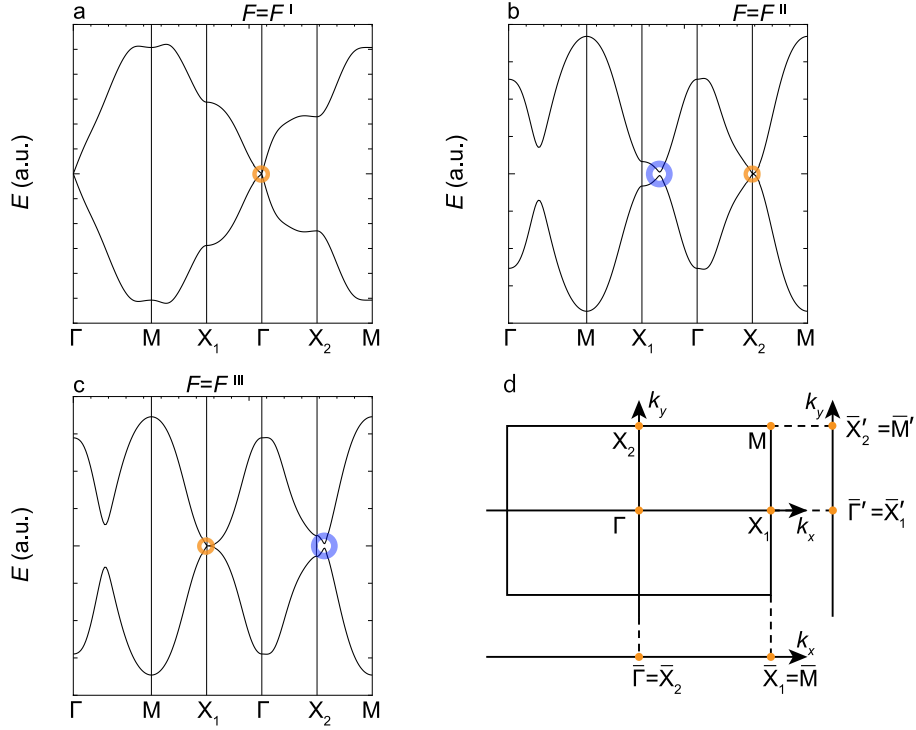}
\caption{{\bf Quasienergy dispersion of 2D-bulk band $E(\boldsymbol{k})$.}
{\bf (a)} Shown is $E(\boldsymbol{k})$ at $F_x^{\rm I}$, where $E_{p(1)}(\boldsymbol{k})$ and $E_{s(-1)}(\boldsymbol{k})$ are degenerate at the $\Gamma$ point (open yellow circle).
{\bf (b)} The same as the panel (a) but at $F_x^{\rm II}$ with the degeneracy at the $X_2$
point (open yellow circle) and the nearly degenerate valleys between the $\Gamma$ and $X_1$ points (open purple circle).
{\bf (c)} The same as the panel (a) but at $F_x^{\rm III}$ with the degeneracy at the 
$X_1$ point open yellow circle) and the nearly degenerate valleys between the $X_2$ and $M$ points (open purple circle).
{\bf (d)} Shown are the high-symmetry points of $\Gamma, X_2, X_1$ and M in the 2D-BZ with their projection onto the $k_x$-axis denoted as $\bar{\Gamma}, \bar{X}_2, \bar{X}_1$ and $\bar{M}$, and onto the $k_y$-axis denoted as $\bar{\Gamma}^\prime, \bar{X}_2^\prime, \bar{X}_1^\prime$ and $\bar{M}^\prime$
respectively.
}
\label{fig3}
\end{center}
\end{figure}

To examine the band crossings in detail, bulk bands $E(\boldsymbol{k})$ at $F_x^{\rm I}$, $F_x^{\rm II}$, and $F_x^{\rm III}$ are shown in Figs.~\ref{fig3}a-\ref{fig3}c, where $E_{p(1)}(\boldsymbol{k})$ and $E_{s(-1)}(\boldsymbol{k})$ are degenerate at a single point of $\boldsymbol{k}^j$ in the 2D-BZ with $j=\Gamma$, $X_2$, and $X_1$, respectively; these are indicated in Fig.~\ref{fig3}d.
Obviously, the crossing points seen in Fig.~\ref{fig2} are found identical with these high-symmetry points projected on the $k_x$-axis, which are denoted as 
$\bar{\Gamma}=\bar{X}_2$ and $\bar{X}_1=\bar{M}$.
Actually, $E(\boldsymbol{k})$ is conical-shaped with linear-dispersion in the vicinity of $\boldsymbol{k}^j$, and this is considered as a DSM state.
It is understood that hereafter, $F_x^I, F_x^{II}$, and $F_x^{III}$ are represented as
$F_x^\Gamma, F_x^{X_2}$, and $F_x^{X_1}$, respectively.
These crossing points are also obtained by inspecting $\mathcal{E}_{p(1)}(k_y)$ and $\mathcal{E}_{s(-1)}(k_y)$ under the boundary condition B.
The high-symmetry points projected on the $k_y$-axis, which are denoted as 
$\bar{\Gamma}^\prime=\bar{X}_1^\prime$ and $\bar{X}_2^\prime=\bar{M^\prime}$, are also depicted in Fig.~\ref{fig3}d.

\bigskip\noindent
{\bf Fourfold accidental degeneracy.}
Here we consider the origin of such band crossings.
Because of the conservation of both T- and pseudo-I-symmetries,
it is still probable that the band crossing between $p(n)$ and $s(n^\prime)$ 
occurs at a high-symmetry point.
In fact, to that end, an additional condition is required that the difference of photon numbers $\Delta n\equiv n-n^\prime$ is an even number.
Contrariwise, when $\Delta n$ is odd, the resulting pair of bands are gapped out; especially, 
the two bands $p(1)$ and $s(0)$ never cross.
Similarly to this case of the $\sigma_z$-conserving interactions, the above results still hold in the case of the $\sigma_z$-non-conserving interactions.
All of the above conditions of band crossings are proved rigorously (see Supplementary Note 3).

\begin{figure}[h]
\begin{center}
\includegraphics[width=16.0cm,clip]{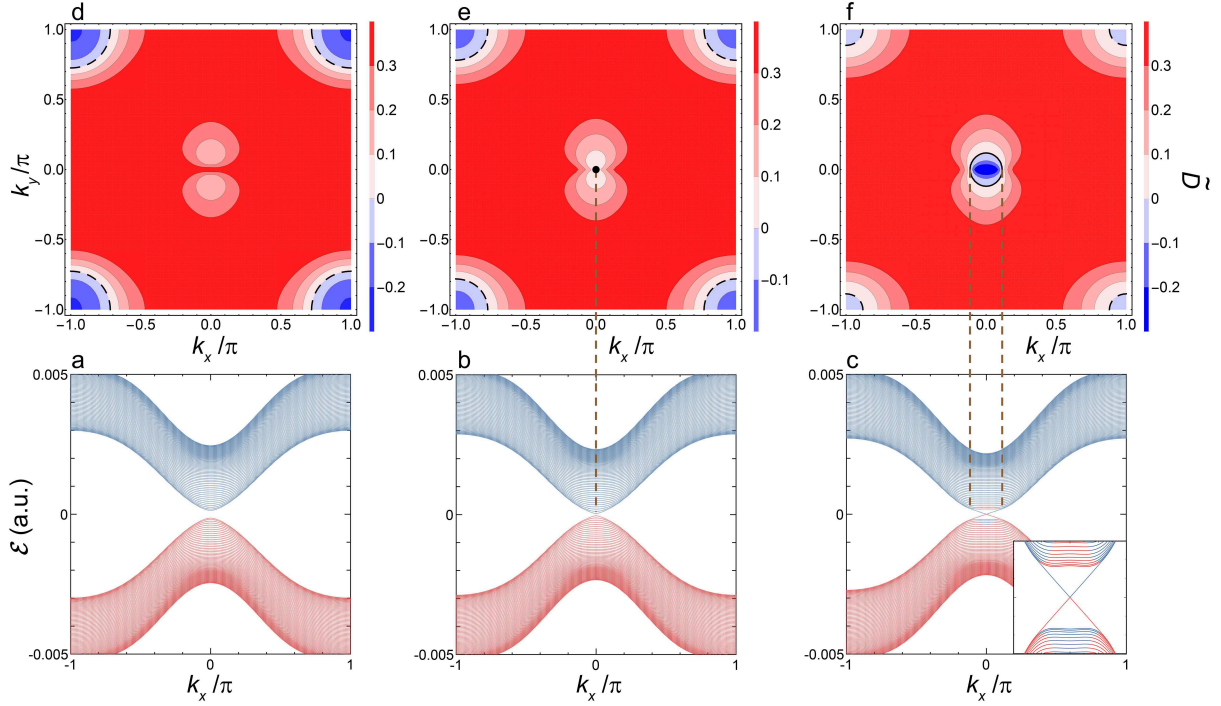}
\caption{{\bf Quasienergy dispersion of $\mathcal{E}(k_x)$ and interband polarization $\tilde{\mathcal{D}}(\boldsymbol{k})$ in the vicinity of $F_x^{\Gamma}$.}
{\bf (a)} Shown are $\mathcal{E}_{p(1)}(k_x)$ and $\mathcal{E}_{s(-1)}(k_x)$
as functions of $k_x$ at $F_x=3.82\times 10^{-4}\: (1.96\: {\rm MV/cm})\: (F_x > F_x^{\Gamma})$.
The two quasienergy bands $p(1)$ and $s(-1)$ are shown by red and blue lines, respectively.
{\bf (b)} The same as the panel (a) but at $F_x^{\Gamma}=3.73\times 10^{-4}\: (1.92\: {\rm MV/cm})$.
{\bf (c)} The same as the panel (a) but at $F_x=3.62\times 10^{-4}\: (1.86\: {\rm MV/cm})\: (F_x^{X_2} < F_x < F_x^{\Gamma})$.
Inset: the expanded view of these two bands in the vicinity of the $\bar{\Gamma}$-point.
{\bf (d)} Shown is a contour map $\tilde{\mathcal{D}}(\boldsymbol{k})$ in the $(k_x,k_y)$-plane
at $F_x $ given by panel (a).
Contours indicating the boundary of $\tilde{\mathcal{D}}(\boldsymbol{k})=0$ are shown by 
black dashed lines.
{\bf (e)} The same as the panel (d) but at $F_x^{\Gamma}$.
Besides, a pinhole indicating $\tilde{\mathcal{D}}(\boldsymbol{k})=0$ at the $\Gamma$-point is 
shown by a black filled circle.
The vertical dashed line shows the projection of $\tilde{\mathcal{D}}(\boldsymbol{k})=0$
(the pinhole) onto the $k_x$-axis shown in the panel (b).
{\bf (f)} The same as the panel (d) but at
$F_x$ given by panel (c). 
Contours indicating the boundary of $\tilde{\mathcal{D}}(\boldsymbol{k})=0$ are shown by 
black solid and dashed lines.
The vertical dashed lines show the projection of $\tilde{\mathcal{D}}(\boldsymbol{k})=0$
(the zero contour) onto the $k_x$-axis shown in the panel (c).
}
\label{fig4}
\end{center}
\end{figure}

\begin{figure}[h]
\begin{center}
\includegraphics[width=16.0cm,clip]{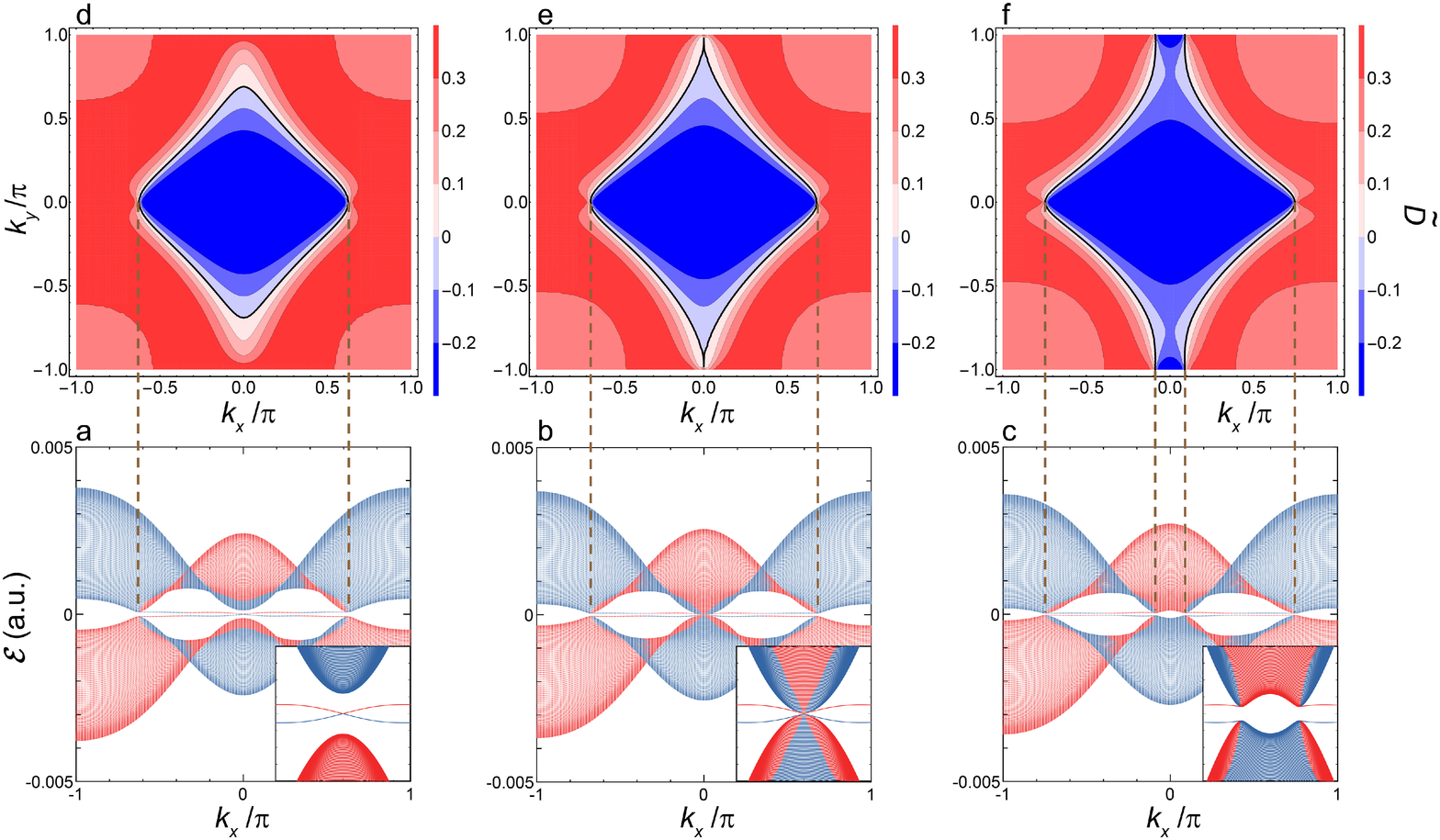}
\caption{{\bf Quasienergy dispersion of $\mathcal{E}(k_x)$ and interband polarization $\tilde{\mathcal{D}}(\boldsymbol{k})$ in the vicinity of $F_x^{X_2}$.}
{\bf (a)} Shown are $\mathcal{E}_{p(1)}(k_x)$ and $\mathcal{E}_{s(-1)}(k_x)$
as functions of $k_x$ at $F_x=2.14\times 10^{-4}\: (1.10\: {\rm MV/cm})\:(F_x^{X_2} < F_x < F_x^\Gamma)$.
The two quasienergy bands $p(1)$ and $s(-1)$ are shown by red and blue lines, respectively.
Inset: the expanded view of these two bands in the vicinity of the $\bar{X}_2$-point.
{\bf (b)} The same as the panel (a) but at $F_x^{X_2}=2.04\times 10^{-4}\: (1.05\: {\rm MV/cm})$.
{\bf (c)} The same as the panel (a) but at $F_x =1.94\times 10^{-4}\: (997\: {\rm kV/cm})\:(F_x^{X_1} < F_x < F_x^{X_2})$.
{\bf (d)} Shown is a contour map $\tilde{\mathcal{D}}(\boldsymbol{k})$ in the $(k_x,k_y)$-plane
at $F_x$ given by panel (a).
The vertical dashed lines show the projection of $\tilde{\mathcal{D}}(\boldsymbol{k})=0$
(the zero contour) onto the $k_x$-axis shown in the panel (a).
{\bf (e)} The same as the panel (d) but at 
$F_x^{X_2}$ and with the zero contour
projected onto the $k_x$-axis shown in the panel (b).
{\bf (f)} The same as the panel (d) but at $F_x$ given by panel (c) and with the zero contour
projected onto the $k_x$-axis shown in the panel (c).
}
\label{fig5}
\end{center}
\end{figure}

\begin{figure}[h]
\begin{center}
\includegraphics[width=16.0cm,clip]{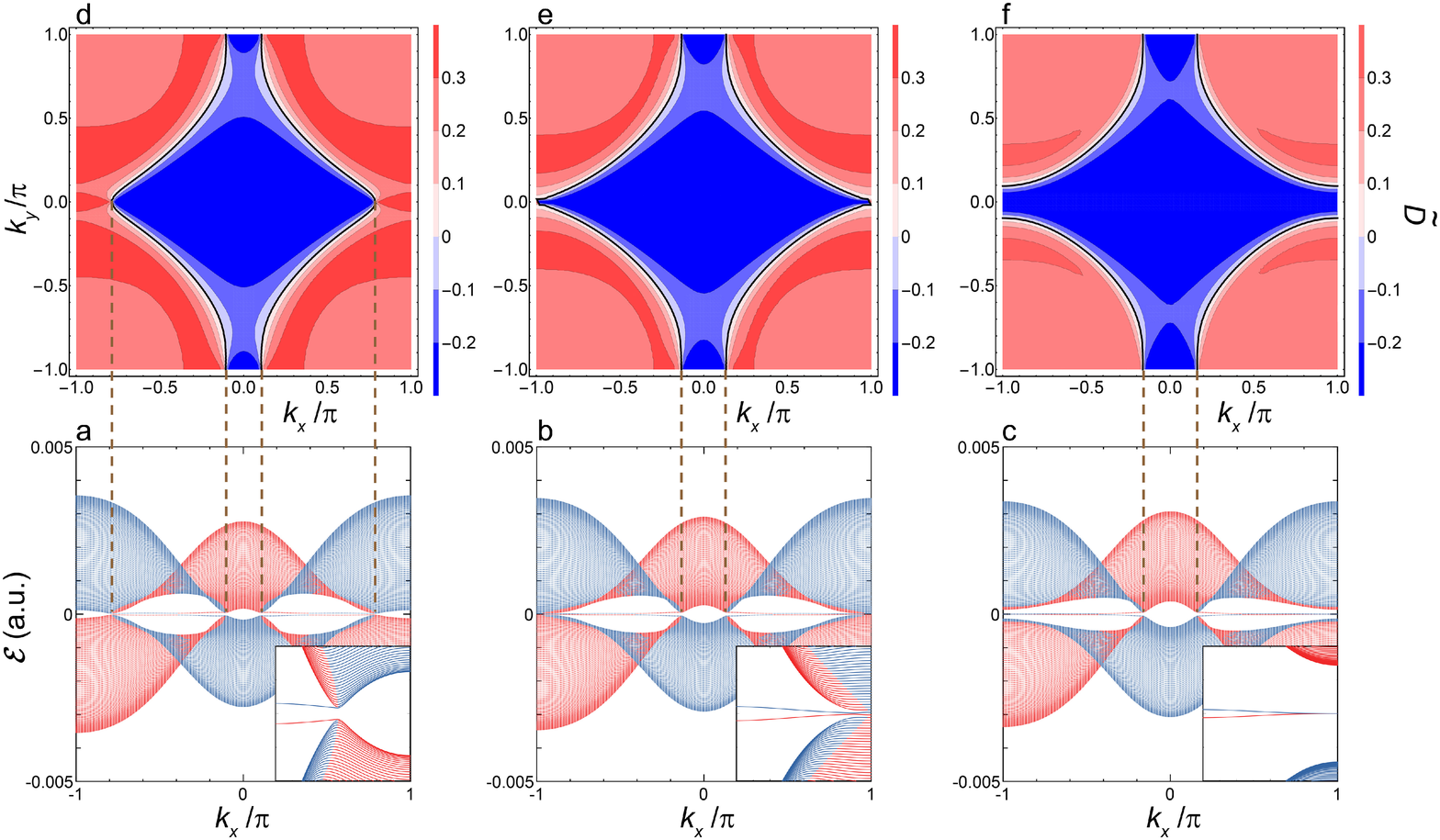}
\caption{{\bf Quasienergy dispersion of $\mathcal{E}(k_x)$ and interband polarization $\tilde{\mathcal{D}}(\boldsymbol{k})$ in the vicinity of $F_x^{X_1}$.}
{\bf (a)} Shown are $\mathcal{E}_{p(1)}(k_x)$ and $\mathcal{E}_{s(-1)}(k_x)$
as functions of $k_x$ at $F_x =1.89\times 10^{-4}\: (974\: {\rm kV/cm})\:(F_x^{X_1} < F_x < F_x^{X_2})$.
The two quasienergy bands $p(1)$ and $s(-1)$ are shown by red and blue lines, respectively.
Inset: the expanded view of these two bands in the vicinity of the $\bar{X}_1$-point.
{\bf (b)} The same as the panel (a) but at $F_x^{X_1} =1.80\times 10^{-4}\: (927\: {\rm kV/cm})$.
{\bf (c)} The same as the panel (a) but at $F_x =1.69\times 10^{-4}\: (871\: {\rm kV/cm})\:(F_x < F_x^{X_1})$.
{\bf (d)} Shown is a contour map $\tilde{\mathcal{D}}(\boldsymbol{k})$ in the $(k_x,k_y)$-plane
at $F_x$ given by panel (a).
The vertical dashed lines show the projection of $\tilde{\mathcal{D}}(\boldsymbol{k})=0$
(the zero contour) onto the $k_x$-axis shown in the panel (a).
{\bf (e)} The same as the panel (d) but at 
$F_x^{X_1}$ and with the zero contour
projected onto the $k_x$-axis shown in the panel (b).
{\bf (f)} The same as the panel (d) but at $F_x$ given by panel (c) and with the zero contour
projected onto the $k_x$-axis shown in the panel (c).
}
\label{fig6}
\end{center}
\end{figure}

\bigskip\noindent
{\bf DSM states and edge states.}
First, we examine the 1D-band, $\mathcal{E}(k_x)$, and concomitant edge states 
obtained under the boundary condition A.
These edge states are either
topologically trivial or non-trivial; hereafter, it is understood that 
the term of the Tamm state\cite{Volkov2016} is used exclusively to mean a trivial state bound on an edge to distinguish it from a non-trivial one. 
Figures~\ref{fig4}a-\ref{fig4}c, \ref{fig5}a-\ref{fig5}c, and \ref{fig6}a-\ref{fig6}c show the spectra of $\mathcal{E}(k_x)$ in the decreasing order of $F_x$.
It is seen that all the DSM states delimits the boundary of a topological phase transition (see Figs.~\ref{fig4}b, ~\ref{fig5}b, and \ref{fig6}b).
It should be noted that the DSM states observed at $F_x^{X_2}$ and $F_x^{X_1}$ coincide with edge states with linear and
nodeful dispersions (see Fig.~\ref{fig5}b, and Fig~\ref{fig6}b, respectively), 
differing from that observed at $F_x^{\Gamma}$ (see Fig.~\ref{fig4}b).
Such edge states are termed the Dirac-Tamm state hereafter just for the sake of convenience of making a distinction from other Tamm states.
As regards the Dirac-Tamm state at $F_x^{X_2}$, with the slight increase of $F_x$ to make a gap open, this is transformed into an unequivocally topological edge state 
with its band structure kept almost as it stands (see Fig.~\ref{fig5}a), while with the change of $F_x$ in the opposite direction, 
this becomes nodeless with two flat dispersions (see Fig.~\ref{fig5}c).
As regards the Dirac-Tamm state at $F_x^{X_1}$, with the slight increase of $F_x$, this is transformed into a nodeless edge state (see Fig.~\ref{fig6}a), while with the slight decrease of $F_x$, this becomes unequivocally topologically trivial (see Fig.~\ref{fig6}c).

\begin{figure}[h]
\begin{center}
\includegraphics[width=16.0cm,clip]{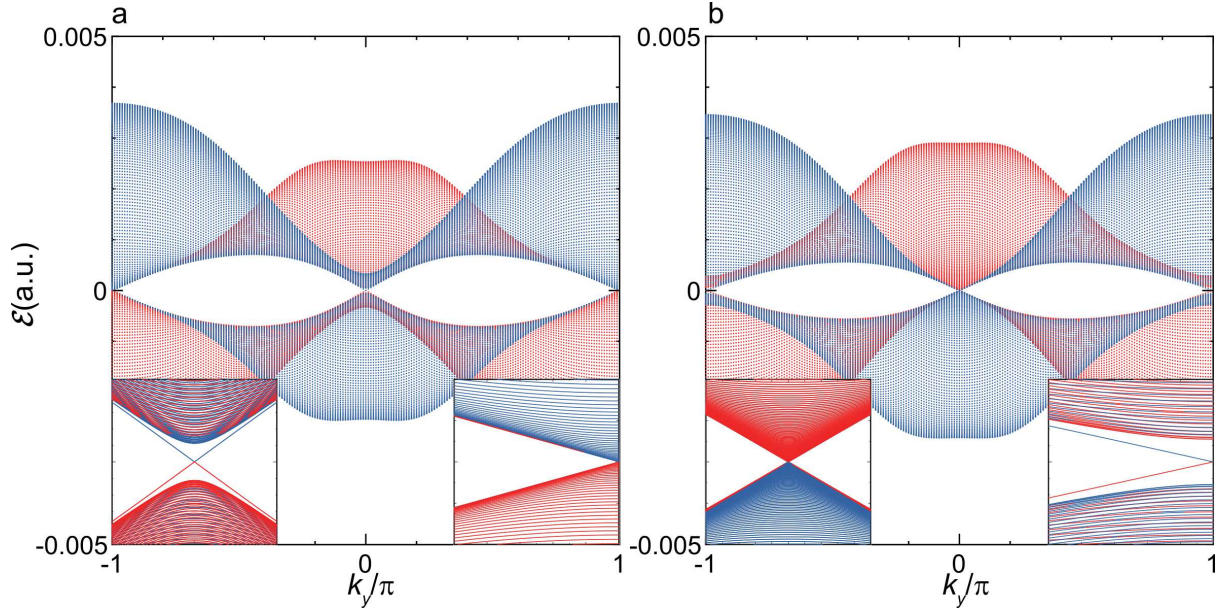}
\caption{{\bf Quasienergy dispersion of $\mathcal{E}(k_y)$ at $F_x^{X_2}$ and $F_x^{X_1}$.}
{\bf (a)} Shown are $\mathcal{E}_{p(1)}(k_y)$ and $\mathcal{E}_{s(-1)}(k_y)$ as functions of $k_y$ at $F_x^{X_2}=2.04\times 10^{-4}\: (1.05\: {\rm MV/cm})$.
The two quasienergy bands $p(1)$ and $s(-1)$ are shown by red and blue lines, respectively.
Insets: the expanded view of these two bands in the vicinity of the $\bar{\Gamma}^\prime (\bar{X}_1^\prime)$-point (left) and the $\bar{X}_2^\prime$-point (right).
{\bf (b)} The same as the panel (a) but at $F_x^{X_1}=1.80\times 10^{-4}\: (927\: {\rm kV/cm})$.
}
\label{fig7}
\end{center}
\end{figure}

Next, we examine the 1D-band, $\mathcal{E}(k_y)$, and concomitant edge states 
obtained under the boundary condition B.
Figure~\ref{fig7} shows the two representative quasienergy bands at $F_x^{X_2}$ and
$F_x^{X_1}$.
Differing from the quasienergy bands shown in Fig.~\ref{fig5}b/Fig.~\ref{fig6}b,
a Dirac-Tamm state is faint and undiscernible
at the $\bar{X}_2^\prime/\bar{X}_1^\prime$ point, 
though a linear and
nodeful dispersion is still discernible around the $\bar{X}_1^\prime/\bar{X}_2^\prime$.
According to these results, it is evident that the specification of the imposed 
boundary condition is crucial for the discussion of the edge states.
Discussion of the origin of such difference will be deepened below.

The topological nature of these edge states is evaluated in terms of the Chern number
of the lower band, denoted as $\alpha_L$, where $E_{\alpha_L}(\boldsymbol{k})\le E_F=0$;
this number is independent of the boundary conditions.
It is confirmed that the non-zero values of $C_{\alpha_L}=1$ are obtained in $F_x^{X_2} < F_x < F_x^{\Gamma}$ and $F_x < F_x^{X_1}$, otherwise this vanishes.
Thus, we verify that the edge state observed in $F_x^{X_1} < F_x < F_x^{X_2}$ is a Tamm state
(see Figs.~\ref{fig5}c and \ref{fig6}a).
Further, the Dirac-Tamm states at $F_x^{X_1}$ and $F_x^{X_2}$ are also considered Tamm states, since their respective net Chern numbers are zero.\cite{Armitage2018}

\bigskip\noindent
{\bf Interband polarization.}
To understand the manifestation of the edge states seen in Fig.~\ref{fig4}-Fig.~\ref{fig7}
 and the boundary-condition dependence,
a macroscopic polarization of the present system, that is,
an induced dipole moment, is examined.
This is given by
\begin{equation}
D_{\boldsymbol{k}\alpha_L}(t)
=\langle\Psi_{\boldsymbol{k}\alpha_L}(t)|x|\Psi_{\boldsymbol{k}\alpha_L}(t)\rangle
=\sum_{bb^\prime(b\not= b^\prime)}[P_{\boldsymbol{k}\alpha_L}(t)]_{bb^\prime}X_{b^\prime b}
\label{D}
\end{equation}
for state $\alpha_L$,
where $x$ is the $x$-component of position vector of electron.
Here, $P_{\boldsymbol{k}\alpha_L}(t)$ represents the associated microscopic interband polarization 
corresponding to an off-diagonal element of a reduced density matrix, and 
$[P_{\boldsymbol{k}\alpha_L}(t)]_{sp}=[P_{\boldsymbol{k}\alpha_L}(t)]_{ps}$
because of $X_{sp}=X_{ps}$.\cite{Haug2009}
The interband polarization in the $\omega$-domain is
introduced as:
\(
\tilde{P}^{(N)}_{\boldsymbol{k}\alpha_L}(\omega)
=(0|D_{\boldsymbol{k}\alpha_L}(t)|N)/X_{sp}
\)
with
\(
\tilde{P}^{(-N)}_{\boldsymbol{k}\alpha_L}(\omega)=
[\tilde{P}^{(N)}_{\boldsymbol{k}\alpha_L}(\omega)]^*
\).
Below, we examine 
$\tilde{\mathcal{D}}(\boldsymbol{k}) \equiv{\rm Re}[\tilde{P}^{(1)}_{\boldsymbol{k}\alpha_L}(\omega)]$
as a function of $\boldsymbol{k}$ in the 2D-BZ; neither
$\tilde{P}^{(N\not= \pm1)}_{\boldsymbol{k}\alpha_L}(\omega)$ nor
${\rm Im}[\tilde{P}^{(\pm 1)}_{\boldsymbol{k}\alpha_L}(\omega)]$
show significant variance in the BZ with the change in $F_x$.
It is stated that 
$\tilde{\mathcal{D}}(\boldsymbol{k})$ precisely reflects the degree of 
parity hybridization in the 2D-BZ that results from 
the I-symmetry breaking.

The calculated results of $\tilde{\mathcal{D}}(\boldsymbol{k})$
are shown in Figs.~\ref{fig4}d-\ref{fig4}f, Figs.~\ref{fig5}d-\ref{fig5}f, and Figs.~\ref{fig6}d-\ref{fig6}f along with $\mathcal{E}(k_x)$
in the vicinity of $F_x^{\Gamma}, F_x^{X_2}$, and $F_x^{X_1}$, respectively,
where a black solid line shows a contour indicating the boundary of $\tilde{\mathcal{D}}(\boldsymbol{k})=0$, which is hereafter termed as the zero contour.
It is readily seen that the zero-contour projected onto
the $k_x$-axis coincides with the segment of the 1D-BZ at which an edge state manifests itself irrespective of being topological or not.
To be more specific,
the edge state is discerned where a vertical line that is parallel to the $k_y$-axis at a certain $k_x$ crosses the zero contour twice.
For instance, as seen in Fig.~\ref{fig6}f, the vertical line crosses this contour twice except around the $\bar{\Gamma}$-point, and the edge state emerges in the corresponding range of $k_x$.
It is remarked that another contour indicating $\tilde{\mathcal{D}}(\boldsymbol{k})=0$ is discerned around the $M$-points in Figs.~\ref{fig4}d-\ref{fig4}f, which is shown by a black dashed line; this causes no edge state and is attributed to an anticrossing between
bands of $s(-1)$ and $p(2)$ where the difference of the respective photon numbers is an odd number.

The above relation of the zero contour with the formation of edge state is straightforward
applied to the case of the boundary condition B.
By projecting $\tilde{\mathcal{D}}(\boldsymbol{k})$'s shown in 
Figs.~\ref{fig4}-\ref{fig6} onto the $k_y$-direction, the existence of edge states and their
forms of manifestation are examined.
According to this, it is speculated that
in the regions of $F_x >F_x^{X_2}$ and $F_x <F_x^{X_1}$, edge states with the similar
patterns to those in the case of the boundary condition A emerge.
In the rest of the regions, edge states exist with the shape of $\infty$ having nodes and antinodes in the whole 1D-BZ.
The various forms of these edge states are schematically depicted in Supplementary Figure 2.
By comparing the 1D-bands shown in Figs.~\ref{fig7}a and \ref{fig7}b with the 
above-speculated results, it is found that
most parts of the $\infty$-shaped edge states are merged into the bulk continuum, and
the Dirac-Tamm states are not discernible around the $\bar{X}_2^\prime$ and $\bar{X}_1^\prime$ points, respectively; though the linear and nodeful dispersions remain just around the $\bar{X}_1^\prime$ and $\bar{X}_2^\prime$ points, respectively.
Therefore, it is stated that the Dirac-Tamm states exist in the case of the boundary condition A, whereas not in the case of the boundary condition B.

\bigskip\noindent
{\bf Nearly fourfold degeneracy and boundary-condition dependence.}
Here we examine the origin of the above-mentioned boundary-condition dependence,
based on the two anticrossings located in between the $X_1$ and $\Gamma$ points
and the $X_2$ and $M$ points seen in Figs.~\ref{fig3}b and \ref{fig3}c, respectively.
The I-symmetry breaking causes band anisotropy of $E(\boldsymbol{k})$, namely, the dependence of band width on the direction of $\boldsymbol{k}$, to form the anticrossing
near the $X_{1(2)}$-point when a gap closes at the $X_{2(1)}$-point.
Even if the gap opens, the anticrossing is sustained in a certain range of $F_x$ with moving in the 2D-BZ, differing from the accidental fourfold degeneracies at the high-symmetry points; these are lifted by slight changes of $F_x$.

Such a property is seen in Figs.~5 and 6, as follows.
The anticrossing along the $X_1-\Gamma$ line is found in $\mathcal{E}(k_x)$'s shown in Figs.~\ref{fig5}a-\ref{fig5}c and Fig.~\ref{fig6}a, and merges into the crossing at the $\bar{X}_1$ point, as shown in Fig.~\ref{fig6}b.
The anticrossing along the $X_2-M$ line is found in $\mathcal{E}(k_x)$'s shown in
Fig.~\ref{fig5}c and Figs.~\ref{fig6}a-\ref{fig6}c, and merges into the crossing at the $\bar{X}_2$ point, as shown in Fig.~\ref{fig5}b.
Thus, these anticrossings are stable against the change of $F_x$ and look nearly fourfold degenerate because of quite small energy separation of the order of 1meV.
Hereafter, for the sake of convenience, very local regions of $\boldsymbol{k}$ 
over which the anticrossings extend along the $X_1-\Gamma$ and $X_2-M$ lines
are termed $V_1$ and $V_2$ points, respectively; further 
the terms of $\bar{V}_1$ and $\bar{V}_2$ points are used as 
the projection onto the $k_x$-direction, respectively.

The singular property around the $V_1$ and $V_2$ points is confirmed 
by seeing the variance of $\tilde{\mathcal{D}}(k_x,0)$ and
$\tilde{\mathcal{D}}(k_x,\pm\pi)$ with respect to $k_x$.
For instance, in Fig.~\ref{fig5}f,
these functions traverse the zero contours with steep changes at the $V_1$ and $V_2$ points, respectively (see Supplementary Figure 3).
Such behavior is attributed to an adiabatic interchange of the constituent of wavefunction $\Psi_{\boldsymbol{k}\alpha_L}(t)$ between $p(1)$ and $s(-1)$ at these points.
This makes $\Psi_{\boldsymbol{k}\alpha_L}(t)$ almost discontinuous, leading to
an abrupt change of parity with the traverse of $k_x$ at these points.
In other words, diabolic-like points are formed at the $V_1$ and $V_2$
points as if monopoles of Berry curvature existed.\cite{Berry1985} 

The existence of the nearly fourfold degeneracies causes more involved
edge-state structure within the gap in $\mathcal{E}_\alpha(k_x)$ than that in $\mathcal{E}_\alpha(k_y)$.
By connecting the points of $\bar{V}_1$ and $\bar{V}_2$ in different manners, 
all of the topological edge states, the Dirac-Tamm states, and Tamm states seen in Figs.~\ref{fig5}a-\ref{fig5}c and Figs.~\ref{fig6}a-\ref{fig6}c are formed in the close vicinity of $\mathcal{E}_\alpha(k_x)=E_F$,
whether there is a Dirac node or not in an edge state.
Besides their topological natures depending on the change of $F_x$,
all these edge states are considered to have the same properties pertinent to the degree of localization of confined electron owing to the same manner of formation; though
the Dirac-Tamm states become delocalized 
just around a local $k_x$-region where bands cross,
due to couplings with continuum of Floquet DSM phases (see Supplementary Note 4).
On the other hand, just topological edge states manifest themselves in $\mathcal{E}_\alpha(k_y)$ without the effect of the nearly fourfold degeneracies at  the points of $V_1$ and $V_2$ (see Fig.~7 and Supplementary Figure 2).
Therefore, it is concluded that the existence of the nearly fourfold degeneracies
is a key effect which governs the manifestation of the Dirac-Tamm states and the Tamm states 
under the boundary condition A.

\bigskip
\noindent
{\bf Discussion}

\noindent
This work shows that the nearly resonant laser-excitation combined with the OSE gives rise to 
the fourfold accidental degeneracies at the high-symmetry points, 
and the resulting Floquet DSM states host unconventional Dirac-Tamm states that are 
transformable into either topological edge states or Tamm states with the change of $F_x$
just under the boundary condition A, differing from the results under the boundary condition B.
A stress is put on the existence of the nearly fourfold degeneracies at the $V_1$ and 
$V_2$ points that arise from the I-symmetry breaking, because these  remain stable against the change of $F_x$ in a certain region of it and fulfills the key role of understanding the different boundary-condition dependence of the edge states.
Such boundary-condition dependence of the present system is reminiscent of graphene
with the zigzag and armchair boundary conditions;\cite{Yao2009, Akhmerov2008, Castro2009}
actually, the Tamm states have almost flat energy dispersion connecting 
the two points of $V_1$ and $V_2$, while graphene has a zigzag edge state with a flat
dispersion  between the $K$ and $K^\prime$ points.

In addition, this study is also related with rapidly noticed studies on the interrelation between a Tamm state and a topological edge state, because the state-of-the-art techniques of fabrication of optical waveguide arrays and photonic crystals have made it possible to create both edge states by mimicking the one-dimensional Su-Schriefer-Hegger model\cite{Longhi2013, Wang2018, Chen2019, Henriques2020, Su1980}
and more complicated systems.\cite{Tsurimaki2018, Lu2019, Latyshev2014}
In this study, both of the edge states are transformed in a continuous manner as a function of the single parameter $F_x$ without changing the composition and structure of the system, which draws a sharp distinction from these existing studies.

Finally, we make comments on the possibility of observing the present findings.
Here, the variance of $F_x$ is around the order of 1MV/cm, leading to high-density electron excitation
with dephasing and population relaxation times of the order of a few hundred fs.
Thus, ultrashort pulse irradiation with $\omega \approx 300$ meV ($T\approx 14$ fs) and 
temporal width of the order of 100 fs is required for realizing band inversion between $E_{s(-1)}(\boldsymbol{k})$ and $E_{p(1)}(\boldsymbol{k})$ to form various types of the edge states.
It would be possible to confirm the manifestation of these states by virtue of the optoelectronic
technique of measuring quasimetallic  photoconductivity produced by  pulse irradiation,
\cite{Auston1975}
which has been utilized for the time-resolved measurement of light-induced Hall effect in graphene. \cite{Mclver2020, Sato2019}
In addition, it is remarked that due to the many-body Coulomb interaction resulting from
intense photoexcitation of electrons, the Floquet bands 
and the values of $F_x$ at which the band inversion and crossing occur
are somewhat modified by the renormalization of carrier energy and Rabi energy.\cite{Haug2009}

\bigskip
\noindent
{\bf Methods}

\noindent
Numerical calculations for a wavefunction $\Psi_{\boldsymbol{k}\alpha}(t)$ of Floquet state $\alpha$ and the associated quasienergy $E_{\alpha}(\boldsymbol{k})$ are implemented by relying on the Fourier-Floquet expansion of Eq.~(\ref{Psi}), followed by diagonalizing the Floquet matrix
\(
\tilde{L}_{nn^\prime}(\boldsymbol{k},\omega)
\).
The explicit expressions of matrix elements of it are given in Supplementary Note 2.
The maximum number of photons $(N_p)$ incorporated in this calculation is three, namely,
$n, n^\prime =-N_p \sim N_p$, and the numerical convergence is checked by using a greater
value of $N_p$.
The following material parameters in the units of a.u. are employed for actual calculations:\cite{Novik2005}
$\epsilon_s=-\epsilon_p=0.01, t_{ss}=t_{pp}=0.001, t_{sp}=0.002, a=12.21$, and $X_{sp}=34.63$.
$\omega$ and $E_g$ are set to be 0.0114 and 0.012, respectively.
Further, the Chern number of a lower band $\alpha_L$ is evaluated by calculating
\begin{equation}
C_{\alpha_L}={1\over 2\pi}\oint d\boldsymbol{k}\cdot \boldsymbol{a}_{\alpha_L}(\boldsymbol{k}),
\label{Chern}
\end{equation}
where the Berry connection is defined by
\(
\boldsymbol{a}_{\alpha_L}(\boldsymbol{k})=-{i\over T}\int^T_0dt\:
\langle \Psi_{\boldsymbol{k}\alpha_L}(t)|\nabla_{\boldsymbol{k}}\Psi_{\boldsymbol{k}\alpha_L}(t)\rangle
\). 

\bigskip

\bigskip\noindent
{\bf Data availability}

\noindent
Data sharing not applicable to this article as no datasets were generated or
analysed during the current study.

\bigskip\noindent
{\bf Code availability}

\noindent
No use of custom code or mathematical algorithm that is deemed central to the conclusions 
of the current study.

\bigskip\noindent
{\bf References}

\vspace*{-1cm}

\bigskip

\bigskip\noindent
{\bf Acknowledgments}

\noindent
This work was supported by JSPS KAKENHI Grant No.~JP19K03695.
The authors are grateful to Prof. J. Fujioka for fruitful discussion.

\bigskip\noindent
{\bf Competing interests}

\noindent
The authors declare no competing financial  or non-financial interests.

\bigskip\noindent
{\bf Author contributions}

\noindent
K.H. conceived the main ideas and supervised the project.
B.Z. carried out the main parts of the numerical calculations, and 
N.M. carried out the rest parts of them. 
All authors discussed and interpreted the results.
K.H. wrote the paper and B.Z. prepared the figures with 
contribution from all authors.

\bigskip\noindent
{\bf Additional information}

\noindent
Supplementary information is available for the paper at
https://xxx.

\newpage


\end{document}